\documentclass[11pt]{article}
\usepackage{moriond2008}
\usepackage{epsfig}
\usepackage[latin1]{inputenc}
\usepackage[OT1]{fontenc}
\newcommand{\journal}[4]
  {{#1}, {\bf #2}, #3 (#4)}
\newcommand{\subjournal}[3]
  {{#1}, #2 (#3)}

\newcommand{\la}
  {\ {\raise-.5ex\hbox{$\buildrel<\over\sim$}}\ }
\newcommand{\ga}
  {\ {\raise-.5ex\hbox{$\buildrel>\over\sim$}}\ }

\newcommand{\unit}[1]
  {{\mbox{\rm\,\,#1}}}
\newcommand{\percent}
  {\,{\rm{per\ cent}}}
\newcommand{\lya}
  {Ly\,$\alpha$}

\newcommand{\micron}
  {{\mbox{$\mu${\rm{m}}}}}

\newcommand{\nir}
  {NIR}

\newcommand{\uprime}
  {\mbox{$u$}}

\newcommand{\iprime}
  {\mbox{$i$}}
\newcommand{\zprime}
  {\mbox{$z$}}

\newcommand{\zmy}
  {\mbox{\zprime$-$$Y$}}
\newcommand{\ymj}
  {\mbox{$Y$$\!\,-$$J$}}

\newcommand{\imy}
  {\mbox{\iprime$-$$Y$}}

\newcommand{\sqdeg}
  {\mbox{${\rm{deg}}^2$}}

\newcommand{\firstsdssqso}
  {SDSS~0836$+$0054}
\newcommand{\secondsdssqso}
  {SDSS~1411$+$1217}

\newcommand{\firstqso}
  {ULAS~J0203$+$0012}
\newcommand{\secondqso}
  {ULAS~J1319$+$0950}

\newcommand{\etal}
  {{\em et al.}}
\newcommand{\ie}
  {{\em i.e.}}
\newcommand{\eg}
  {{\em e.g.}}

\newcommand{\ApJ}
  {ApJ}

\newcommand{\MNRAS}
  {MNRAS}
\newcommand{\AaA}
  {A\&A}
\newcommand{\AJ}
  {AJ}
\newcommand{\ApJS}
  {ApJS}

\newcommand{\PASP}
  {PASP}

\newcommand{\probqso}
  {P_{\rm{q}}}

\begin{document}
\vspace*{4cm}
\title{THE UKIRT INFRARED DEEP SKY SURVEY\\
  AND THE SEARCH FOR THE MOST DISTANT QUASARS}

\author{ DANIEL J.\ MORTLOCK, MITESH PATEL, STEPHEN J.\ WARREN,}
\address{Astrophysics Group, Blackett Laboratory,
  Imperial College London, Prince Consort Road,\\
  London SW7~2AZ, United Kingdom}

\author{ BRAM P.\ VENEMANS, RICHARD G.\ McMAHON, PAUL C.\ HEWETT,}
\address{Institute of Astronomy, 
  Madingley Road, Cambridge CB3~0HA, United Kingdom}

\author{ CHRIS SIMPSON,}
\address{Astrophysics Research Institute, Liverpool John Moores University, 
  Egerton Wharf,\\ Birkenhead CH41~1LD, United Kingdom}

\author{ ROB G.\ SHARP}
\address{Anglo-Australian Observatory, PO~Box~296, Epping, NSW~1710, Australia}

\maketitle\abstracts{The UKIRT Infrared Deep Sky Survey (UKIDSS) 
Large Area Survey (LAS) has the necessary combination of filters 
($Y$, $J$, $H$ and $K$), depth ($Y \la 20.2$) and 
area coverage ($\sim 4000$\,\sqdeg)
to detect several redshift $z \ga 6.4$ quasars.
The Third Data Release (DR3) included 
$\sim 1000$ \sqdeg\ of LAS observations which have so far yielded
two previously known $z \simeq 6$ quasars 
and two new discoveries:  
\firstqso, at $z = 5.72$;
and 
\secondqso, at $z = 6.13$.}


\noindent
High-redshift quasars are unique probes of the early Universe 
because they are the only non-transient sources which are 
sufficiently luminous that high signal--to--noise ratio spectra
can routinely be obtained (\eg, Schneider 1999).
Such observations not only reveal their intrinsic properties
(\eg, Walter \etal\ 2004)
but also
probe the intervening matter via absorption.
The most striking demonstration of this has come from 
studies of 
redshift $z \simeq 6$ quasars 
(Fan \etal\ 2001; Willott \etal\ 2007),
which have revealed 
a sharp increase in the \lya\ optical depth beyond $z \simeq 5.7$
(\eg, Becker \etal\ 2001).
Combined with the 
results from the
{\em{Wilkinson Microwave Anisotropy Probe}}
cosmic microwave background (CMB) measurements 
(\eg, Dunkley \etal\ 2008),
these quasar observations 
contradict
most simple ionization histories 
(\eg, Gnedin 2000),
leaving such intriguing possibilities as 
double reionization (\eg, Furlanetto \& Loeb 2005).

Further progress in understanding the
ionization history of the Universe will require 
the discovery of the first quasars at $z \simeq 7$.
The most distant quasars known at present 
(\eg, 
CFHQS~J2329$-$0301, at $z = 6.43$, Willott \etal\ 2007;
SDSS~1148$+$5251, at $z = 6.42$, Fan \etal\ 2003)
have been found by looking for point--sources with
very red optical colours in wide-field surveys
like the Sloan Digital Sky Survey (SDSS; York \etal\ 2000)
and the Canada France High-$z$ Quasar Survey (CFHQS; Willott \etal\ 2007),
but optical searches are unlikely to probe beyond the current 
redshift limits
due to an unfortunate combination of astrophysics and detector technology.
On the one hand, almost all $z \simeq 6$ photons with wavelengths
shorter than the \lya\ transition at $\lambda = 0.1216$\,\micron\
are absorbed by intervening hydrogen, and so sources are effectively
dark below $\lambda \simeq [0.85 + 0.12 (z - 6)]$\,\micron.
Conversely, most optical charge-coupled device (CCD) 
detectors have a poor
response beyond wavelengths of $\lambda \simeq 0.9$\,\micron\
(\ie, redward of the \zprime\ or $Z$ bands),
so quasars at $z \ga 6.4$ are destined to remain 
invisible to CCD-based surveys.
Combined with the low numbers of $z \ga 6$ quasars 
(\eg, a surface density of $\sim 0.02 \unit{deg}^{-2}$; Jiang \etal\ 2007)
progress can only be made with wide-field surveys at longer wavelengths,
most obviously in the near-infrared (\nir).
The largest completed \nir\ survey, 
the 2 Micron All Sky Survey (2MASS; Skrutskie \etal\ 2006),
only reaches $J \simeq 15.8$,
and 
does not have sufficient depth to find any plausible high-redshift 
quasars; hence there has been a great need for deeper wide-field \nir\
surveys.
The 
Visible and Infrared Survey Telescope for Astronomy 
(VISTA; Emerson \etal\ 2004) should 
cover $\sim 2 \times 10^4$\,\sqdeg\ to $J \simeq 20$ during the next decade,
but the most immediate progress in the search for $z \simeq 7$ quasars
will come from the UKIRT Infrared Deep Sky Survey (UKIDSS).

UKIDSS (Lawrence \etal\ 2007) is a suite of five \nir\ surveys being undertaken 
using the Wide Field Camera (WFCAM; Casali \etal\ 2007)
on the United Kingdom Infrared Telescope (UKIRT).
One of these,
the Large Area Survey (LAS),
was designed to have sufficient area coverage 
($\sim 4000$\,\sqdeg) and depth 
(detection of point--sources with a signal--to--noise ratio of 
$5$ at $Y \simeq 20.2$) 
to find several $z \ga 6.4$ quasars.
Its footprint is matched to that of SDSS,
and so in the UKIDSS LAS area there will
exist complementary imaging covering the wavelength range from 
$\sim 0.35$\,\micron\ (the SDSS \uprime\ band)
to 
$\sim 2.4$\,\micron\ (the UKIDSS $K$\ band).

\begin{figure}
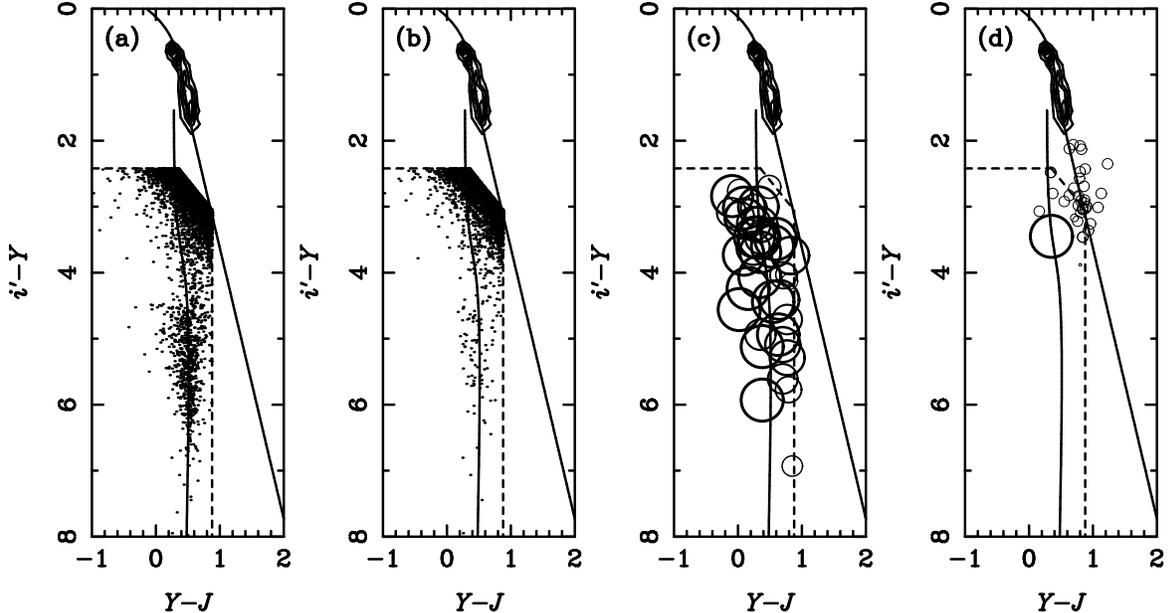

\psfig{figure=iyj_unfiltered.eps,height=8.1cm}
\psfig{figure=iyj_filtered.eps,height=8.1cm}
\psfig{figure=iyj_original.eps,height=8.1cm}
\psfig{figure=iyj_reobserved.eps,height=8.1cm}
\caption{Colour--colour diagrams of a subset of UKIDSS quasar candidates,
  before filtering (a), 
  after filtering (b),
  selected according to quasar probability, $\probqso$ (c),
  and after follow-up photometry (d).
  A sample of bright Main Sequence stars
  (contours) is also shown,
  along with fiducial star and quasar loci (solid lines)
  and the initial colour cut (dashed lines).
  In (c) and (d) the symbol size scales with $\probqso$.
  The large symbol in (d) 
  (\ie, the only candidate with $\probqso \simeq 1$ after follow-up) is
  the $z = 6.13$ quasar \secondqso\ (Mortlock \etal\ 2008b).}
\label{figure:iyj}
\end{figure}

Another important aspect in the design of UKIDSS
is the use of the 
newly developed $Y$ band filter,
which lies between the \zprime\ (or $Z$) and $J$ bands
and has
significant response in the wavelength range
$0.97\,\micron \la \lambda \la 1.07\,\micron$
(Warren \& Hewett 2002; Hillenbrand \etal\ 2002; Hewett \etal\ 2006).
Not only will all quasars with $z \la 7.2$ have significant 
emission over the whole of the $Y$ band, 
but they are expected to be bluer in $\ymj$ than 
the L and T dwarfs with which they would otherwise be confused
(\eg, Warren \& Hewett 2002).

UKIDSS observations began in 2005, 
and there have been a total of $\sim 10^6$ 
science exposures as of December 2007.
The data are made available in incremental releases,
first to
European Southern Observatory (ESO) countries
and then, 18 months later, to the world,
via the 
WFCAM Science Archive\footnote{The WSA is located at 
  {\tt{http://surveys.roe.ac.uk/wsa/}}.} 
(WSA; Hambly \etal\ 2008).
The Third Data Release (DR3),
made available in December 2007 to ESO,
includes $\sim 1000$\,\sqdeg\ of LAS
imaging in (at least) the $Y$ and $J$ bands.

\begin{figure}
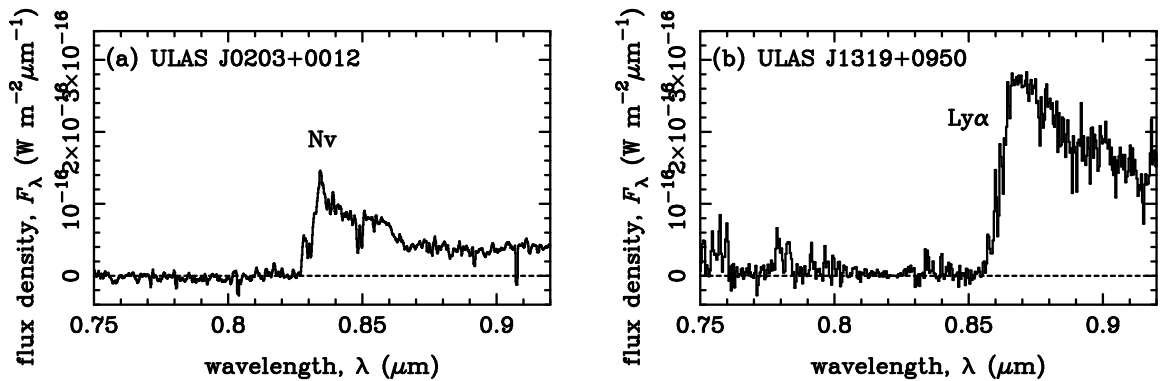

\psfig{figure=firstqso.eps,height=5.5cm}
\psfig{figure=secondqso.eps,height=5.5cm}
\caption{Spectra of the two new high-redshift quasars
  discovered so far in UKIDSS:
  (a) \firstqso, at $z = 5.72$ 
  (Venemans \etal\ 2007),
  and
  (b) \secondqso, at $z = 6.13$ 
  (Mortlock \etal\ 2008b).}
\label{figure:quasars}
\end{figure}

There should be several $z \simeq 6$ quasars
amongst the 
$\sim 5 \times 10^7$ sources catalogued in the DR3 LAS.
Applying the obvious astronomical criteria 
(that high-redshift quasars are expected to be seen as 
point--sources which are very red in $\imy$ or $\zmy$ and blue in $\ymj$)
immediately removes $\sim 99 \percent$ of sources 
from consideration but, in DR3, still leaves $\sim 10^5$ 
``pre-candidates'', as shown in Fig.~\ref{figure:iyj}~(a).
Most of these are ``glitches'' of 
either the data acquistition or subsequent processing,
but in many cases the underlying cause 
can be identified and accounted for,
finally leaving the sample of predominantly real sources seen in 
Fig.~\ref{figure:iyj}~(b).

Critically, this sample is generated by 
a fairly well understood statistical process: 
aside from any actual high-$z$ quasars present,
these sources are M, L and T dwarfs randomly scattered to have quasar-like 
colours.
The observational noise can be modelled,
as can the star and quasar populations,
which means that it is possible to calculate the relative likelihoods
that a member of each of these two populations would be 
measured to have the colours of any given candidate (Mortlock \etal\ 2008a).
Folding in the relative numbers of quasars and stars 
(\ie, that there are far more of the latter) 
then gives the probability, $\probqso$, that each candidate is a quasar,
and in Fig.~\ref{figure:iyj}~(c) the small number of sources
with $\probqso \geq 0.01$ are shown.
Importantly, the calculation of $\probqso$ is based on 
fluxes, rather than flux ratios,
so two sources with identical colours can actually
have quite different values of $\probqso$,
which is particularly relevant close to the survey limit.
Because all
the available information is included in the calculation of $\probqso$,
all the candidates can be compared objectively,
with the only significant limitation being the degree to which it is 
possible to model the extremes of the observational error distributions.

Having calculated $\probqso$ for each candidate, they can be ranked 
and the most promising sources selected for 
follow-up photometry,
with
\iprime\ band observations
especially effective because most candidates 
are so close to the SDSS \iprime\ limit that any deeper measurement
provides significant extra information
(Mortlock \etal\ 2008a).
The utility of follow-up imaging
(as opposed to spectroscopy)
can be seen by comparing Fig.~\ref{figure:iyj}~(c) and (d),
which shows that 
routine photometric observations are sufficient to reveal that
most candidates are just scattered M dwarfs.

Fortunately, follow-up photometry does not reject all the candidates,
and to date UKIDSS DR3 has yielded four high-$z$ quasars.
These include 
the successful recovery of 
\firstsdssqso\ (Fan \etal\ 2001)
and 
\secondsdssqso\ (Fan \etal\ 2004),
as well as two new discoveries:
\firstqso\ (Venemans \etal\ 2007),
at\footnote{The initial redshift estimate of $z = 5.86$ was revised 
to $z = 5.72$ after \firstqso\ was found to be broad absorption line
quasar (Mortlock \etal\ 2008b).} $z = 5.72 \pm 0.01$
and with $Y = 19.9 \pm 0.1$
(\ie, at the limit of detectability in UKIDSS);
and 
\secondqso\ (Mortlock \etal\ 2008b),
at $z = 6.13 \pm 0.01$ 
and with $Y = 19.10 \pm 0.03$.
Spectra of both are shown in Fig.~\ref{figure:quasars}.

The identification of four $z \simeq 6$ quasars in the UKIDSS DR3 dataset
is consistent with expectations from the Fan \etal\ (2001) 
quasar luminosity function,
and thus 
represents a complete end--to--end verification of the survey.
Aside from showing that 
the UKIDSS data are of the necessary quality,
it also validates 
the cross-matching to the SDSS and 2MASS catalogues,
and demonstrates that it is possible to produce a manageable candidate sample
using almost completely automated procedures.
Whilst it is unlikely that 
the first $z \simeq 7$ quasar
is amongst the remaining 
UKIDSS DR3 candidates,
the above successes give reason for 
confidence that several $z \ga 6.4$ quasars 
will be in the increasingly complete LAS.
More such discoveries will come in the next decade as 
VISTA and various longer wavelength surveys begin to make observations,
and the detection of $z \simeq 7$ quasars may even become routine, 
but for the moment they represent the absolute limit of observational
astronomy.


\section*{Acknowledgments}

The ambitious project would not have been possible without the work 
of the hundreds of engineers and scientists 
who made the SDSS and UKIDSS projects the great successes they are.



\end{document}